\documentclass[11pt,a4paper]{article}
\usepackage{jinstpub}	
\pdfoutput=1 	

\usepackage{graphicx} 	
\usepackage{amssymb} 	
\usepackage{mathtools} 	
\usepackage{comment} 	
\usepackage{color} 		
\usepackage{multirow}	
\usepackage[]{siunitx}  
\DeclareSIUnit\sample{S}
\DeclareSIUnit\bits{bits}

\hypersetup{
	pdfinfo ={
    	Author={C.G. Davis et. al.},
        Title={De-noising for EXO-200},
        Subject={EXO-200}}}
\usepackage{xspace} 	
\usepackage[all]{hypcap} 

\DeclarePairedDelimiter\ExpectFences{[}{]} 
\newcommand{\Expect}[1]{\operatorname{\mathbb{E}}\ExpectFences*{#1}}
\DeclarePairedDelimiterX\ExpectCondFences[2]{[}{]}{#1\,\delimsize\vert\,\mathopen{}#2}
\newcommand{\ExpectCond}[2]{\operatorname{\mathbb{E}}\ExpectCondFences*{#1}{#2}}
\DeclarePairedDelimiter\VarFences{[}{]}
\newcommand{\Var}[1]{\operatorname{var}\VarFences*{#1}}
\DeclarePairedDelimiterX\VarCondFences[2]{[}{]}{#1\,\delimsize\vert\,\mathopen{}#2}

\DeclarePairedDelimiterX\CovarFences[2]{[}{]}{#1, #2}
\newcommand{\Covar}[2]{\operatorname{cov}\CovarFences*{#1}{#2}}
\DeclarePairedDelimiterX\CovarCondFences[3]{[}{]}{#1, #2\,\delimsize\vert\,\mathopen{}#3}
\newcommand{\CovarCond}[3]{\operatorname{cov}\CovarCondFences*{#1}{#2}{#3}}

\newcommand{\znbb} {$0\nu\!\beta\!\beta$\xspace}
\newcommand{\tnbb} {$2\nu\!\beta\!\beta$\xspace}
\newcommand{\otsx} {$^{136}\mathrm{Xe}$\xspace}

\newcommand{\sep}{, } 

\makeatletter
\renewcommand\tableofcontents{%
    \@starttoc{toc}%
}
\makeatother

\begin{document}

\title{An Optimal Energy Estimator
to Reduce Correlated Noise for the EXO-200 Light Readout}
\collaboration{EXO-200}

\newcommand{\Alabama}{p}
\newcommand{\Bern}{n}

\newcommand{\Carleton}{r}
\newcommand{\CSU}{k}
\newcommand{\Drexel}{q}
\newcommand{\Duke}{c}
\newcommand{\IBS}{v}
\newcommand{\IHEP}{j}
\newcommand{\Illinois}{d}
\newcommand{\Indiana}{b}
\newcommand{\ITEP}{e}
\newcommand{\Laurentian}{l}
\newcommand{\SNOLAB}{m}
\newcommand{\Maryland}{a}
\newcommand{\McGill}{g}
\newcommand{\Munich}{s}
\newcommand{\SDakota}{w}
\newcommand{\SLAC}{f}
\newcommand{\Stanford}{o}
\newcommand{\Stony}{u}
\newcommand{\TRIUMF}{h}
\newcommand{\UMass}{t}
\newcommand{\WIPP}{x}

\author[\Maryland,1]{C.G.~Davis\note{Corresponding author}}
\author[\Maryland]{C.~Hall}

\author[\Indiana]{J.B.~Albert}
\author[\Duke]{P.S.~Barbeau}
\author[\Illinois]{D.~Beck}
\author[\ITEP]{V.~Belov}
\author[\SLAC]{M.~Breidenbach}
\author[\McGill,\TRIUMF]{T.~Brunner}
\author[\ITEP]{A.~Burenkov}
\author[\IHEP]{G.F.~Cao}
\author[\IHEP]{W.R.~Cen}
\author[\CSU]{C.~Chambers}
\author[\Laurentian,\SNOLAB]{B.~Cleveland}
\author[\Illinois]{M.~Coon}
\author[\CSU]{A.~Craycraft}
\author[\SLAC]{T.~Daniels}
\author[\ITEP]{M.~Danilov}
\author[\Indiana]{S.J.~Daugherty}
\author[\SLAC]{J.~Davis}
\author[\Bern]{S.~Delaquis}
\author[\Laurentian]{A.~Der Mesrobian-Kabakian}
\author[\Stanford]{R.~DeVoe}
\author[\Alabama]{T.~Didberidze}
\author[\TRIUMF]{J.~Dilling}
\author[\ITEP]{A.~Dolgolenko}
\author[\Drexel]{M.J.~Dolinski}
\author[\Carleton]{M.~Dunford}
\author[\CSU]{W.~Fairbank Jr.}
\author[\Laurentian]{J.~Farine}
\author[\Munich]{W.~Feldmeier}
\author[\UMass]{S.~Feyzbakhsh}
\author[\Munich]{P.~Fierlinger}
\author[\Stanford,1]{D.~Fudenberg}
\author[\Carleton,\TRIUMF]{R.~Gornea}
\author[\Carleton]{K.~Graham}
\author[\Stanford]{G.~Gratta}
\author[\Alabama]{M.~Hughes}
\author[\Stanford]{M.J.~Jewell}
\author[\SLAC]{A.~Johnson}
\author[\Indiana,2]{T.N.~Johnson\note{Now at University of California, Davis, California 95616, USA}}
\author[\UMass]{S.~Johnston}
\author[\ITEP]{A.~Karelin}
\author[\Indiana]{L.J.~Kaufman}
\author[\Carleton]{R.~Killick}
\author[\Carleton]{T.~Koffas}
\author[\Stanford]{S.~Kravitz}
\author[\TRIUMF]{R.~Kr\"{u}cken}
\author[\ITEP]{A.~Kuchenkov}
\author[\Stony]{K.S.~Kumar}
\author[\IBS]{D.S.~Leonard}
\author[\Carleton]{C.~Licciardi}
\author[\Drexel]{Y.H.~Lin}
\author[\Illinois,3]{J.~Ling\note{Now at Sun Yat-Sen University, Guangzhou, China}}
\author[\SDakota]{R.~MacLellan}
\author[\Munich]{M.G.~Marino}
\author[\SLAC]{B.~Mong}
\author[\Stanford]{D.~Moore}
\author[\Stony]{O.~Njoya}
\author[\WIPP]{R.~Nelson}
\author[\SLAC]{A.~Odian}
\author[\Stanford]{I.~Ostrovskiy}
\author[\Alabama]{A.~Piepke}
\author[\UMass]{A.~Pocar}
\author[\SLAC]{C.Y.~Prescott}
\author[\TRIUMF]{F.~Reti\`{e}re}
\author[\SLAC]{P.C.~Rowson}
\author[\SLAC]{J.J.~Russell}
\author[\Stanford]{A.~Schubert}
\author[\Carleton,\TRIUMF]{D.~Sinclair}
\author[\Drexel]{E.~Smith}
\author[\ITEP]{V.~Stekhanov}
\author[\Stony]{M.~Tarka}
\author[\IHEP]{T.~Tolba}
\author[\Alabama]{R.~Tsang}
\author[\Stanford]{K.~Twelker}

\author[\Bern]{J.-L.~Vuilleumier}
\author[\SLAC]{A.~Waite}
\author[\Illinois]{J.~Walton}
\author[\CSU]{T.~Walton}
\author[\Stanford]{M.~Weber}
\author[\IHEP]{L.J.~Wen}
\author[\Laurentian]{U.~Wichoski}
\author[\WIPP]{J.~Wood}
\author[\Illinois]{L.~Yang}
\author[\Drexel]{Y.-R.~Yen}
\author[\ITEP]{O.Ya.~Zeldovich}

\affiliation[\Maryland]{Physics Department, University of Maryland, College Park, Maryland 20742, USA}
\affiliation[\Indiana]{Physics Department and CEEM, Indiana University, Bloomington, Indiana 47405, USA}
\affiliation[\Duke]{Department of Physics, Duke University, and Triangle Universities Nuclear Laboratory (TUNL), Durham, North Carolina 27708, USA}
\affiliation[\Illinois]{Physics Department, University of Illinois, Urbana-Champaign, Illinois 61801, USA}
\affiliation[\ITEP]{Institute for Theoretical and Experimental Physics, Moscow, Russia}
\affiliation[\SLAC]{SLAC National Accelerator Laboratory, Menlo Park, California 94025, USA}
\affiliation[\McGill]{Physics Department, McGill University, Montr\'eal, Qu\'ebec, Canada}
\affiliation[\TRIUMF]{TRIUMF, Vancouver, British Columbia, Canada}
\affiliation[\IHEP]{Institute of High Energy Physics, Beijing, China}
\affiliation[\CSU]{Physics Department, Colorado State University, Fort Collins, Colorado 80523, USA}
\affiliation[\Laurentian]{Department of Physics, Laurentian University, Sudbury, Ontario P3E 2C6, Canada}
\affiliation[\SNOLAB]{SNOLAB, Sudbury, Ontario, Canada}
\affiliation[\Bern]{LHEP, Albert Einstein Center, University of Bern, Bern, Switzerland}
\affiliation[\Stanford]{Physics Department, Stanford University, Stanford, California 94305, USA}
\affiliation[\Alabama]{Department of Physics and Astronomy, University of Alabama, Tuscaloosa, Alabama 35487, USA}
\affiliation[\Drexel]{Department of Physics, Drexel University, Philadelphia, Pennsylvania 19104, USA}
\affiliation[\Carleton]{Physics Department, Carleton University, Ottawa, Ontario K1S 5B6, Canada}
\affiliation[\Munich]{Technische Universit\"at M\"unchen, Physikdepartment and Excellence Cluster Universe, Garching 80805, Germany}
\affiliation[\UMass]{Amherst Center for Fundamental Interactions and Physics Department, University of Massachusetts, Amherst, MA 01003, USA}
\affiliation[\Stony]{Department of Physics and Astronomy, Stony Brook University, SUNY, Stony Brook, New York 11794, USA}
\affiliation[\IBS]{IBS Center for Underground Physics, Daejeon, Korea}
\affiliation[\SDakota]{Physics Department, University of South Dakota, Vermillion, South Dakota 57069, USA}
\affiliation[\WIPP]{Waste Isolation Pilot Plant, Carlsbad, New Mexico 88220, USA}

\emailAdd{claytongdavis@gmail.com}
\emailAdd{dfudenb@stanford.edu}
\arxivnumber{1605.06552}
\date{\today}

\abstract{%
The energy resolution of the EXO-200 detector is limited
by electronics noise in the measurement of the scintillation response.
Here we present a new technique to extract optimal scintillation energy measurements
for signals split across multiple channels
in the presence of correlated noise.
The implementation of these techniques improves
the energy resolution of the detector 
at the neutrinoless double beta decay Q-value from 
\mbox{$\left[1.9641\pm 0.0039\right]$\%} to 
\mbox{$\left[1.5820\pm 0.0044\right]$\%}.}


\maketitle


\section{Introduction}
\label{sec-intro}

Searches for neutrinoless double beta decay (\znbb) are the 
most sensitive test of
whether neutrinos are Majorana particles%
~\cite{
Giuliani:2012zu,
Cremonesi:2013vla,
Barabash:2014bfa,
Dell'Oro:2016dbc}. 
In the simplest interpretation,
if \znbb decay occurs, 
the emitted electron pair must carry away the full energy (Q-value) of the decay.
In contrast,
for the observed two neutrino decay (\tnbb)~\cite{Agashe:2014kda},
the electron pair takes on a continuum of lesser energies.
The existence of 
\znbb decay would be observed as a peak in the energy spectrum at the Q-value 
broadened by the energy resolution of the detector.
Thus, 
obtaining the best energy resolution possible
for a given detection technology
is important to maximize the sensitivity of
searches for \znbb decay.

EXO-200 searches for the \znbb decay of $^{136}$Xe
in liquid xenon (LXe) time projection chambers~\cite{Albert:2014awa}.
EXO-200 determines the energy of interactions in the
LXe (events) by combining the ionization and scintillation response.
This allows energy resolution to be improved by accounting for
the anti-correlation between these responses~\cite{Conti:2003av}.
Currently,
noise in the electronic readout of the scintillation response
limits
the energy resolution of EXO-200.
Detectors collecting scintillation light produced by an event
are read out as multiple channels.
The existing readout electronics introduce noise
that is correlated between these channels.
In early EXO-200 analyses, 
the scintillation signal per event was determined
from the collective response of all light readout channels.
This enhanced the correlated noise components,
leading to degradation of the resolution.

Correlated noise is a common problem in detectors
with many signal readout channels,
including in those of other \znbb experiments.
Previous techniques implemented to reduce correlated noise
have taken advantage of the ability to measure
cross-correlations between the noise in the signal channel
and correlated channels where only noise
is present~\cite{%
ManciniTerracciano:2012fq,
Ouellet:2015}. 

Here we present a technique
capable of accounting for 
both correlated noise and simultaneous signals
on all scintillation channels in EXO-200.
This technique, known as de-noising,
was developed for the \znbb analysis reported in \cite{Albert:2014awa} 
and used in subsequent data analyses.
As described in detail in sections~\ref{sec-algorithm}~and~\ref{sec-characterization},
this technique 
does not produce filtered waveforms.
Instead, it takes advantage of features of
the scintillation collection, measurement, and readout process in EXO-200
to create an optimal energy estimator 
that is robust against the correlated noise shared between channels.
Similar non-filtering techniques for optimally estimating pulse magnitudes
have been considered for single channels~\cite{Bryant:2010}.%

The benefits of such improved energy estimates are
observed when considering the entire EXO-200 data set,
and are discussed in section~\ref{sec-res-improve}.
The general method presented can be adapted
to a variety of detector technologies 
via changes to the detector model.

\section{EXO-200 Detector Description}\label{sec-detector}

\begin{figure}[t]
    \centering
    \includegraphics%
[width=\columnwidth]
{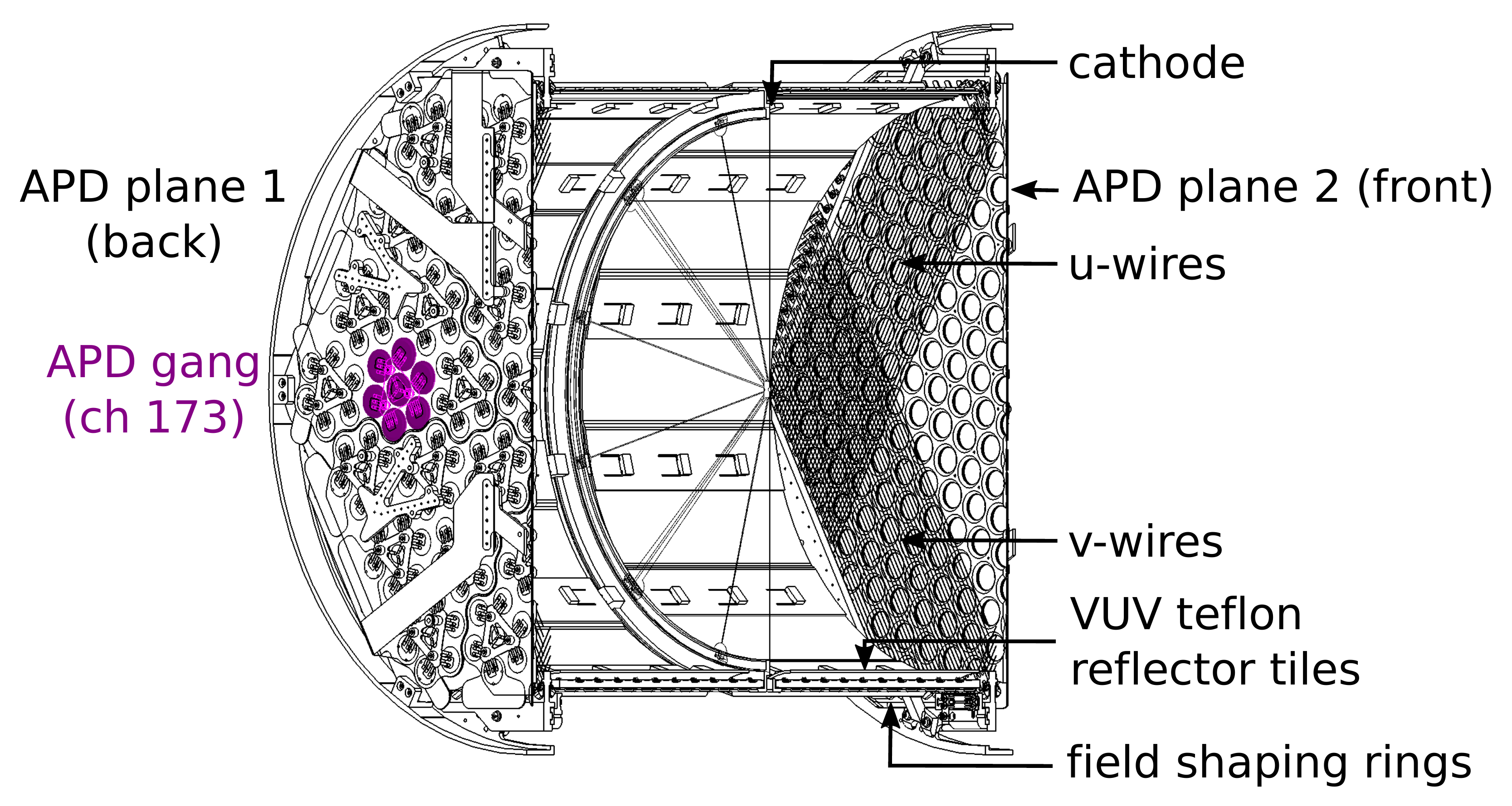}
    \caption{%
A schematic showing a cutaway of the EXO-200 detector inside of the LXe vessel
with relevant components identified.
The sensitive side of an APD plane is visible to the right behind the u- and v-wires used for the ionization signal,
showing individual circular APDs.
At the left, the backside of an APD plane is visible, 
and the back of the channel 173 APD gang is highlighted in magenta. 
Teflon reflectors are placed around the interior of the field shaping rings.}
    \label{fig-detector}
\end{figure}

Here we describe the EXO-200 detector,
focusing on features of 
interest to the measurement of the scintillation response;
for a more complete description see~\cite{Auger:2012gs}.
The LXe vessel in EXO-200 is 
\SI{40}{\cm} in diameter and
\SI{44}{\cm} in length 
and contains 
LXe enriched to 80.6\% in \otsx 
(Q-value of $Q_{\beta\beta}{=}2457.83 \pm 0.37~\si{\kilo\eV}$ \cite{Redshaw:2007un}).
This vessel is split into two cylindrical time projection chambers (TPCs)
by a shared cathode,
as shown in figure~\ref{fig-detector}.
Both TPCs are instrumented with readout systems to
measure induced and collected charge
and scintillation light signals.

The isotropically emitted \SI{178}{\nano\m} scintillation light
produced by energy deposits in the LXe
is collected in each TPC behind the anode
by an array of large area avalanche photodiodes (APDs)~\cite{Neilson:2009kf}.
The shared cathode has 
$\sim$90\% optical transparency~\cite{Auger:2012gs},
so every energy deposit
can generate scintillation signals on all APDs.
Such a signal is a pulse 
proportional to the amount of light collected by that APD.
Light collection efficiency is improved
by Teflon (PTFE) reflector tiles
along the length of the detector
just inside the field shaping rings.
Each array of APDs (APD plane) consists of 234 unencapsulated APDs~\cite{our:apds} 
mounted on a copper platter that
provides a common APD cathode bias voltage of \SI{-1400}{\volt}.
Neighboring APDs,
with similar gain characteristics,
are `ganged' in groups of 5--7.
Each gang is read out as a single channel
by a charge-sensitive preamplifier.
An anode trim voltage,
adjustable on groups of six channels by up to \SI{100}{\volt},
allows APD gains to be matched 
to within 2.5\% of a nominal gain of 200.

The signals from each APD gang are shaped using 
two RC integrators, each with a time constant, $\tau$, of \SI{3}{\micro\s}, 
and two CR differentiators, each with $\tau$ of \SI{10}{\micro\s}.
An additional differentiator with $\tau$ of \SI{300}{\micro\s}
is included in the preamplifier circuit.
These shaping times are much longer 
than the light collection and APD response time,
resulting in an APD signal with a fixed pulse shape
that is fully determined by this shaping.
All the channels are digitized at \SI{1}{\mega\sample\per\s}
and saved 
in `frames' of 2048 sequential samples
centered around a readout trigger.  

Each APD plane also includes a PTFE diffuser that can be illuminated
by a \SI{405}{\nano\m} laser pulser 
for monitoring of the APD light response.
Further monitoring
is performed using a $^{228}$Th calibration source
with a prominent $\gamma$ line at \SI{2615}{\keV}
(from the decay of $^{208}$Tl).
This source is deployed,
within a guide tube positioned around the exterior of the LXe vessel,
at several locations
to fully illuminate the LXe volume.
Data from such deployments are used to determine 
the position and time dependence of the scintillation response, 
which is described in section~\ref{sec-lightmap}.

\subsection{APD Noise}
\label{sec-apdnoise}
\begin{figure}[t]
    \centering
	\includegraphics[width=\columnwidth]{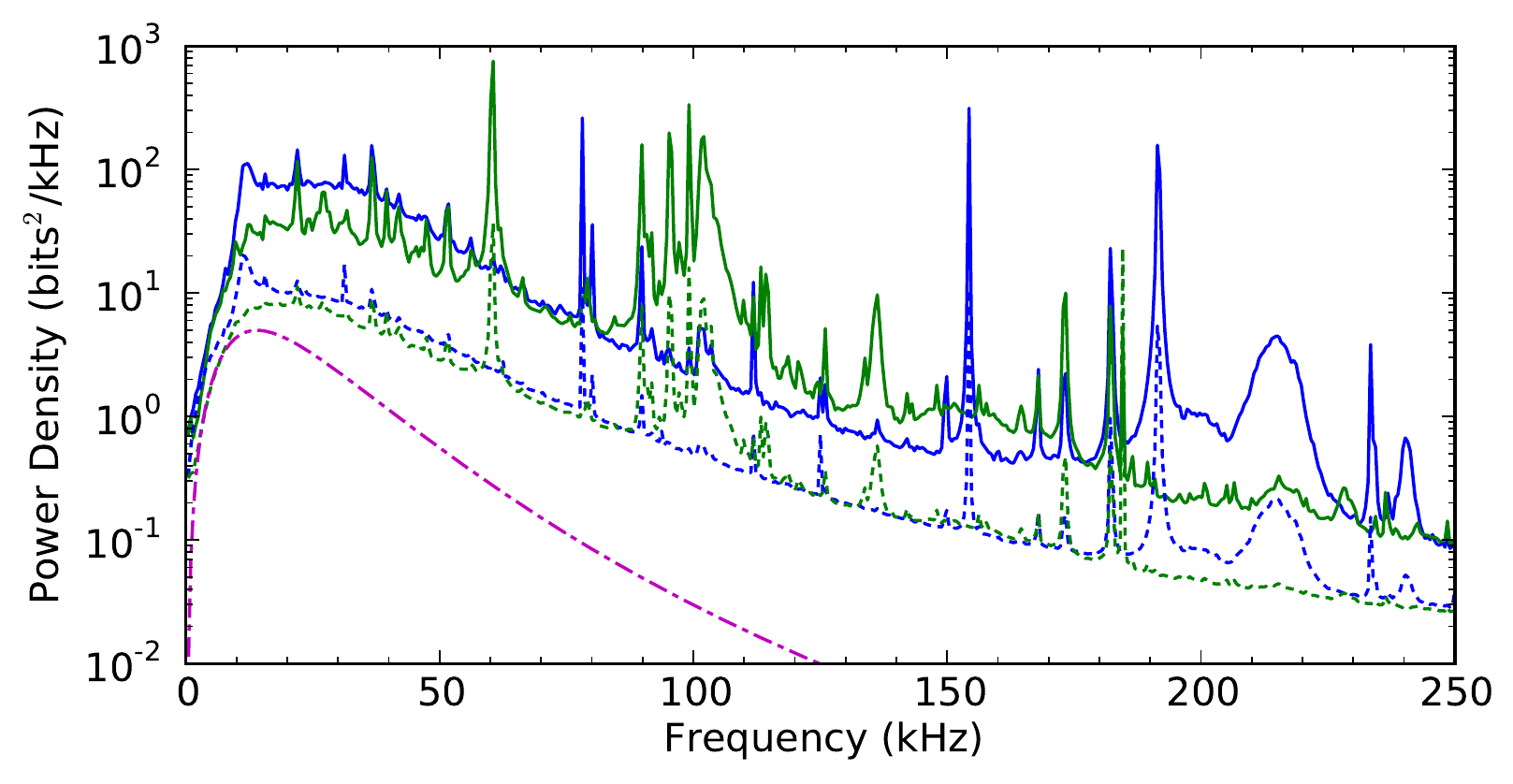}
\caption{%
Coherent (solid) and incoherent (dotted) sums of
noise power spectral density for each APD plane,
TPC 1 (blue) and TPC 2 (green),
from a noise run in January 2014.
The APD pulse power spectrum with arbitrary normalization (magenta dash-dotted line) is also shown.}
	\label{fig-enoise}
\end{figure}
\begin{figure}[t]
\centering
\setlength\tabcolsep{3pt}
\begin{tabular}{m{0.90\columnwidth}m{0.09\columnwidth}}
\vspace{0pt}\includegraphics[width=0.90\columnwidth, trim = {0 0 55bp 0}, clip]{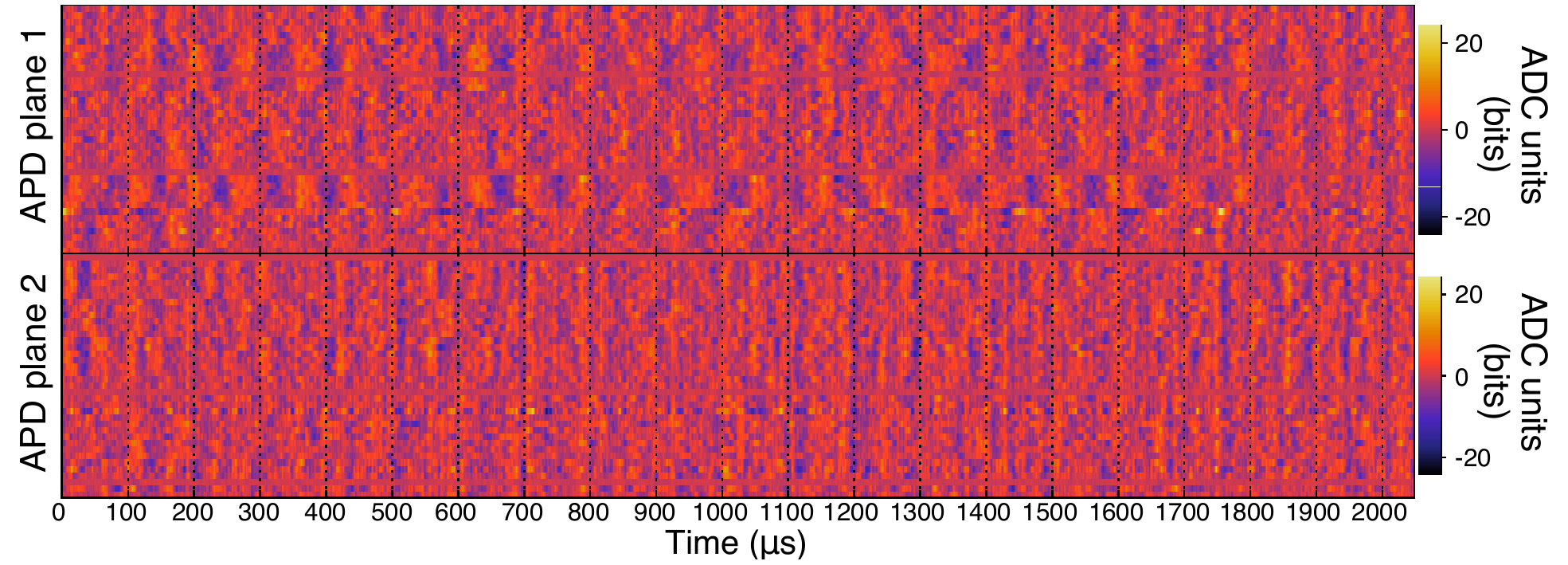}&
\vspace{0pt}\includegraphics[width=0.09\columnwidth, trim = {513px  0 0 89bp}, clip]{005800-APDs--.pdf}
\end{tabular}
\caption{%
APD waveforms from a single noise event.
Each APD channel is shown as separate vertical bin.
The baseline subtracted magnitude for
each sampling time is shown.
Vertical features are indicative of correlations in the noise across channels.}
\label{fig-noiseapdchannels}
\end{figure}

The power spectrum of the pulse shape is compared in figure~\ref{fig-enoise} to power spectra of the noise.
For each APD plane, the noise power spectrum of 
the summed waveform over all APD channels (the coherent sum)
is compared to 
the sum of individual channel power spectra (the incoherent sum).
The large excess of noise in the coherent sum indicates 
that the noise is dominated 
by sources that are correlated between channels.
This correlated nature of the noise can be seen in figure~\ref{fig-noiseapdchannels} as vertical features of common intensity at a point in time.
APD gangs in a given anode trim group are more correlated and appear as blocks. 

Exhaustive study of the detector electronics indicates
that this correlated noise largely results from 
the grounding scheme 
and switching noise from the high voltage supply of the APD plane.
This noise will be reduced in future data collection
subsequent to an upcoming electronics upgrade.

The development of data analysis techniques
that account for such correlated noise 
allows substantial improvement 
in the scintillation energy resolution
that can be obtained with existing EXO-200 data.
Section~\ref{sec-algorithm} explains 
the de-noising technique utilized,
section~\ref{sec-characterization} describes the 
characterization of EXO-200 required to implement the de-noising,
and section~\ref{sec-res-improve} discusses the improvement obtained.

\section{Optimal Energy Estimator} 
\label{sec-algorithm}

In EXO-200, 
uncertainty in the scintillation energy measurement is 
understood to have two primary sources:
the detector electronics contribute a substantial additive noise
to the readout channels,
and the photon yield from typical events is sufficiently low 
that individual channels are subject to significant shot noise 
(statistical fluctuation in readout quanta).
De-noising is the pursuit of an energy estimator 
that is robust to both of these forms of noise,
while only demanding a level of detector characterization 
which is reasonable to achieve in EXO-200.

Partial solutions to this problem are unsatisfactory in practice.
As described in section~\ref{sec-apdnoise}, 
a simple sum over all channels,
which decreases shot noise relative to 
the magnitude of the summed signal pulse,
does not suppress the correlated components of 
the additive noise.
A potential solution to this problem could be 
to take the sum of channels weighted
to suppress the correlated noise.
If EXO-200 had numerous signal-free channels
this would be a reasonable solution,
but in practice,
the channels with the most highly correlated noise
also have correlated signals,
leading to cancellation of both signal and noise.

Qualitatively, 
a full solution must be capable of using a priori knowledge of
correlations between channels due to noise correlations
and 
correlations between channels of signal pulses for each event.
These known properties of the signal pulses depend on 
timing and the position of energy deposits
and therefore vary for each event.
For each event the method should use this knowledge to select channels,
and frequencies within those channels,
which reduce shot noise without introducing too much additive noise.

The remainder of this section describes a de-noising technique
that balances additive noise against shot noise
in a multichannel detector
to produce a minimum-variance unbiased linear estimator of energy.
As only linear estimators are considered,
the optimization is of a quadratic function of several variables subject to linear equality constraints.
This optimization is of a standard form and can be solved efficiently using highly
optimized linear algebra software.
As the technique is applicable to a variety of detector technologies,
details of the input parameters specific to EXO-200
are deferred to section~\ref{sec-characterization}.
The required form of and
constraints on the estimator is derived
for frames containing only a single event (eq.~\ref{eq-OptCriterion3}).
For a more thorough treatment 
readers are directed to~\cite{Clayton:2014}.
The signal and noise are treated in detail
to reduce the obtained estimator solutions 
to eq.~\ref{eq-LagrangeSystemOfEquations}.
These solutions are further reduced 
to eq.~\ref{eq-LagrangeSystemOfEquations-reduced} 
for the purpose of computation. 
The extension of this technique to frames with 
more complex events is then discussed.

\subsection{Optimization Framework}

For each event, 
a frame of waveform data is collected
(see section~\ref{sec-detector}).
Once the timing of the scintillation signal is determined,
the waveform observed on channel $i$, $X_i$,
can be written in the (discrete) frequency domain with complex-values as
\begin{equation}\label{eq-WaveformModel1}
X_i[f] = M_i Y_i[f] + N_i[f],
\end{equation}
where $Y_i$ is the known (unit magnitude) pulse shape,
$M_i$ is the pulse magnitude,
$N_i$ is the additive noise on the channel, 
and $f$ is the frequency index.
In this notation, 
$M_i$ is a random variable that depends on the scintillation energy, 
and $N_i[f]$ is a waveform of complex-valued random variables 
that are independent of the scintillation energy.

We seek an estimator $\widehat{E}$
for the scintillation energy $E$ of the form:%
\begin{equation}\label{eq-EstimatorForm}
\widehat{E} = 
\sum_{i,f} \left( A_i[f] X_i^{\rm R}[f] + B_i[f] X_i^{\rm I}[f] \right),
\end{equation}
where R (I) denotes the real (imaginary) part
of the complex value, 
and $A_i[f]$ and $B_i[f]$ are real-valued weighting parameters
to be derived on an event-by-event basis
so that eq.~\ref{eq-EstimatorForm}
produces the optimal energy estimate.   
This estimator must be unbiased:
\begin{equation}\label{eq-Constraint}
E = \Expect{\widehat{E}} =
\sum_{i, f} \Bigg( A_i[f] \Expect{X_i^{\rm R}[f]} + B_i[f] \Expect{X_i^{\rm I}[f]} \Bigg),
\end{equation}
where $\mathbb{E}$ returns the expectation value.
We seek the minimum-variance estimator
subject to this constraint,
where the variance is expressed as%
\begin{equation}\label{eq-OptCriterion1}
\Var{\widehat{E}}{=}
\sum_{i, j, f, g}\left(
\begin{multlined}
A_i[f]A_j[g] \Covar{X_i^{\rm R}[f]}{X_j^{\rm R}[g]} {+} A_i[f]B_j[g] \Covar{X_i^{\rm R}[f]}{X_j^{\rm I}[g]}\\
+ B_i[f]A_j[g] \Covar{X_i^{\rm I}[f]}{X_j^{\rm R}[g]} {+} B_i[f]B_j[g] \Covar{X_i^{\rm I}[f]}{X_j^{\rm I}[g]}
\end{multlined}
\right).
\end{equation}

Assuming that the expectation values and pairwise covariances
of waveform samples can be specified,
eq.~\ref{eq-Constraint} and eq.~\ref{eq-OptCriterion1} form
a quadratic optimization problem with a linear constraint 
over the parameters $A_i[f]$, $B_i[f]$.
Using the method of Lagrange multipliers,
we minimize $\Var{\widehat{E}} + \lambda (\Expect{\widehat{E}} - E)$ for some $\lambda$.
The optima are described by solutions to the linear systems,
for all $i$ and $f$, of equations:%
\begin{subequations}\label{eq-OptCriterion3}\begin{align}
\sum_{j,g} A_j[g] \Covar{X_i^{\rm R}[f]}{X_j^{\rm R}[g]} + B_j[g] \Covar{X_i^{\rm R}[f]}{X_j^{\rm I}[g]} 
&= \lambda \Expect{X_i^{\rm R}[f]},\\
\sum_{j,g} A_j[g] \Covar{X_i^{\rm I \;}[f]}{X_j^{\rm R}[g]} + B_j[g] \Covar{X_i^{\rm I \;}[f]}{X_j^{\rm I}[g]}
&= \lambda \Expect{X_i^{\rm I}[f]},\\
\sum_{j,g} A_j[g] \Expect{X_j^{\rm R}[g]} + B_j[g] \Expect{X_j^{\rm I}[g]} 
&= E.
\end{align}\end{subequations}

The expectation values and covariances of 
$X_i^{\rm R}[f]$ and $X_i^{\rm I}[f]$ depend on
the position and energy of an event, 
as well as the date
(due to time-varying gain and noise properties of the EXO-200 detector).
Given these dependencies and 
the large number of pairwise covariances which must be specified,
detailed characterization of the EXO-200 waveforms is a challenging task 
and is deferred to section~\ref{sec-characterization}.
Section~\ref{sec-exp-val} describes simplifying assumptions
applicable to a broad range of detectors.

\subsection{Waveform Expectation Values}
\label{sec-exp-val}

This section describes generic assumptions
on the statistical model of the waveform, $X_i[f]$
(see eq.~\ref{eq-WaveformModel1}).
The additive noise waveforms at each frequency, $N_i[f]$, are assumed to have mean zero:%
\begin{subequations}\begin{align}%
\Expect{N_i[f]} 
&= 0\text{\quad for all $i$, $f$,}\\
\intertext{and to be independent from the pulse magnitude:}
\Covar{N_i^{\rm R}[f]}{M_j} = \Covar{N_i^{\rm I \;}[f]}{M_j} 
& = 0\text{\quad for all $i$, $j$, $f$.}
\end{align}\end{subequations}
These assumptions are valid at the energies of interest for \znbb,
where the signals are sufficiently large that the trigger time is independent of the noise.
The energy dependence of the expectation value
of the signal pulse magnitude
is assumed linear:%
\begin{equation}\label{eq-lightmapdefinition}
\Expect{M_i} = L_i(\vec{x}, t) E,
\end{equation}
where $L_i(\vec{x}, t)$,
the light map (see section~\ref{sec-lightmap}),
depends on the APD channel~$i$,
the position of the energy deposit $\vec{x}$,
and the date of the event $t$.
In reality, 
there is some non-linear relationship between
the energy of a deposit
and the number of photons generated,
which is determined from calibrations.
This approach still works for deviations from linearity
that are common to all channels, 
but differing non-linearity across channels is not addressed.
For non-zero frequencies we assert 
that additive noise waveforms ($N_i[f]$)
at different frequencies must have zero correlation:%
\begin{equation}\label{eq-noisecovariszero}
\Covar{N_i^{\rm I}[f]}{N_j^{\rm I}[g]} =
\Covar{N_i^{\rm R}[f]}{N_j^{\rm I}[g]} =
\Covar{N_i^{\rm R}[f]}{N_j^{\rm R}[g]} =
0\text{\quad for $f {\ne} g$ and all $i$,$j$.}
\end{equation}
It is possible that noise at different frequencies is related by higher-order statistics.
However, 
assuming that the phases of the noise components are independent of the trigger time,
eq.~\ref{eq-noisecovariszero} will still hold.

Given these assumptions,
eqs.~\ref{eq-OptCriterion3} reduce (see \cite{Clayton:2014}) to:
\begin{subequations}\label{eq-LagrangeSystemOfEquations}
\begin{gather}\allowdisplaybreaks 
\begin{multlined}[b]
\sum_j \left(\Covar{N_i^{\rm R}[f]}{N_j^{\rm R}[f]} A_j[f] + \Covar{N_i^{\rm R}[f]}{N_j^{\rm I}[f]} B_j[f]\right)\\
{}+\sum_{j,g} \Covar{M_i}{M_j} Y_i^{\rm R}[f]Y_j^{\rm R}[g] A_j[g] + \sum_{j,g} \Covar{M_i}{M_j} Y_i^{\rm R}[f]Y_j^{\rm I}[g] B_j[g]\\
= \lambda L_i(\vec{x}, t) Y_i^{\rm R}[f] \text{\quad for all $i,f$},%
\end{multlined}\\
\begin{multlined}[b]
\sum_j \left(\Covar{N_i^{\rm I}[f]}{N_j^{\rm R}[f]} A_j[f] + \Covar{N_i^{\rm I}[f]}{N_j^{\rm I}[f]} B_j[f]\right) \\
{}+\sum_{j,g} \Covar{M_i}{M_j} Y_i^{\rm I}[f]Y_j^{\rm R}[g] A_j[g] + \sum_{j,g} \Covar{M_i}{M_j} Y_i^{\rm I}[f]Y_j^{\rm I}[g] B_j[g]\\
= \lambda L_i(\vec{x}, t) Y_i^{\rm I}[f] \text{\quad for all $i,f$},%
\end{multlined}\\
\sum_j \left(\sum_{g} A_j[g] Y_j^{\rm R}[g] + B_j[g] Y_j^{\rm I}[g]\right) L_j\left(\vec{x}, t\right) = 1.\label{eq-LagrangeSystemOfEquations-c}
\end{gather}\end{subequations}%
For any particular event,
eqs.~\ref{eq-LagrangeSystemOfEquations} comprise a well-defined 
inhomogenous system of equations.
This system is linear in the unknown parameters
($A_i[f]$, $B_i[f]$, and $\lambda$),
and quadratic in the waveforms (via the covariances of the signals).
The true event energy does not explicitly appear although 
it does enter into these covariances of the signals.
This apparent circularity is addressed 
in section~\ref{subsec-ScintillationNoiseModel}.
Analysis of the additive noise,
light map,
and shot noise
$\left(\Covar{N_i[f]}{N_j[f]}, L_i(\vec{x}, t),\text{ and }\Covar{M_i}{M_j}\right)$
is deferred to section~\ref{sec-characterization}.

\subsection{Computational Implementation}
\label{sec-comp-imp}

In principle, 
from eqs.~\ref{eq-LagrangeSystemOfEquations}
it is straightforward to form the equivalent matrix expression
${\bf A}y  = c$,
where $y$ contains the unknown parameters, 
and to apply 
standard matrix algorithms.%
\footnote{As performed in section 4.6 of~\cite{Clayton:2014}.}
In practice, 
EXO-200 has 1024 frequency values and $\sim$70 scintillation channels,
so eqs.~\ref{eq-LagrangeSystemOfEquations} describe a linear system in ${\sim}1.4\times 10^5$ (real) unknowns.
As this system must be solved separately for each
of more than $10^8$ EXO-200 events,
the direct matrix approach is computationally infeasible.

Fortunately,
the equivalent matrix operator ${\bf A}$
can be applied to vectors efficiently.
The matrix operator can be computed readily 
after first evaluating the contribution to the energy estimate of a unit collection on each channel:%
\begin{equation}
p_j \coloneqq \sum_f Y^{\rm R}_j[f] A_j[f] + Y^{\rm I}_j[f] B_j[f] 
\text{\quad for all $j$.}
\end{equation}
Eqs.~\ref{eq-LagrangeSystemOfEquations} then reduce to:
\begin{subequations}\label{eq-LagrangeSystemOfEquations-reduced}
\begin{gather}\allowdisplaybreaks 
\begin{multlined}[b]
\sum_j \left(\Covar{N_i^{\rm R}[f]}{N_j^{\rm R}[f]} A_j[f] + \Covar{N_i^{\rm R}[f]}{N_j^{\rm I}[f]} B_j[f]\right)
+\sum_{j} \Covar{M_i}{M_j} Y_i^{\rm R}[f] p_j \\
= \lambda L_i(\vec{x}, t) Y_i^{\rm R}[f] \text{\quad for all $i,f$},
\end{multlined}\\
\begin{multlined}[b]
\sum_j \left(\Covar{N_i^{\rm I}[f]}{N_j^{\rm R}[f]} A_j[f] + \Covar{N_i^{\rm I}[f]}{N_j^{\rm I}[f]} B_j[f]\right)
+ \sum_{j} \Covar{M_i}{M_j} Y_i^{\rm I}[f] p_j \\
= \lambda L_i(\vec{x}, t) Y_i^{\rm I}[f] \text{\quad for all $i,f$},
\end{multlined}\\
\sum_j p_j L_j\left(\vec{x}, t\right) = 1.
\end{gather}\end{subequations}%
Although the number of unknowns is not reduced,
the number of products required to evaluate ${\bf A}y$ is only $\sim$$0.1\%$
of the number for a dense matrix of similar size.
This system can then be solved using any iterative method
designed to take advantage of fast matrix-vector multiplication. 
EXO-200 uses the biconjugate gradient stabilized 
algorithm~\cite{BlBiCGSTAB}.
In addition,
the problem is also parallelizable,
as each event can be de-noised independently.

\subsection{Extensions for Complex Events}
\label{sec-complex-events}

The preceding sections describe EXO-200's approach
to estimate the scintillation energy of simple frames,
where exactly one event deposits energy at a single site (SS) within the detector.
Candidate \znbb and \tnbb events primarily fall in this category.
The EXO-200 data set also includes frames capturing multiple events and events depositing energy at multiple sites.
Most of the backgrounds produce multisite (MS) events,
where multiple charge deposits are resolved 
but all scintillation light is detected simultaneously.
In this section,
we discuss approaches to de-noise these two kinds of complex frames.

First, consider frames with scintillation signals from multiple events
that are separated by some measurable amount of time.
Such frames are infrequent during EXO-200 data taking,%
\footnote{Except for time-correlated backgrounds 
such as the decays of ${}^{214}$Bi followed by that of ${}^{214}$Po
used in~\cite{Albert:2015vma}
and the prompt cosmogenic events
used in~\cite{EXO200::2015wtc}.}
but common in calibration runs due to the high event rate of the radioactive sources.
Discarding such triggers from calibration data leads to an
unacceptable loss of live time.  
Instead, 
the energy estimators are additionally constrained
by placing nulls between them.
In other words,
we wish to constrain the energy estimator for each event 
to be unaffected by the signal pulses from any other event.
The constraint of eq.~\ref{eq-LagrangeSystemOfEquations-c} is generalized as%
\begin{equation}\label{eq-LagrangeSystemOfEquations-MultiEventConstraint}
\sum_j \left(\sum_{g} A_{ja}[g] Y_{jb}^{\rm R}[g] + B_{ja}[g] Y_{jb}^{\rm I}[g]\right) L_j\left(\vec{x}_b, t\right) 
= \delta_{ab}\text{\quad for all $a$, $b$,}
\end{equation}
where $A_j[f]$, $B_j[f]$, $Y_j[f]$ and $\vec{x}$ now depend on
the event contained within the waveform as indexed by $a, b$.
We are taking advantage of the different timing of the waveforms through the phases of $Y_{a}[f]$,
and the different distribution of
photons across channels through the light map's dependence on $\vec{x}$, 
to differentiate signal pulses.
As the events become closer in time and space,
this constraint becomes harder to satisfy
and the resolution of the energy estimator degrades accordingly.
However, 
the errors are uncorrelated between events sharing a waveform 
(to the extent that the light map, waveforms, and position estimates are accurate).

For the second type of complex frame, those with MS events,
the de-noising analysis of the preceding sections is confronted
with multiple positions $\vec{x}_1$,\ldots $\vec{x}_n$.
Currently, EXO-200 only considers the total scintillation energy of the MS event,
by extending the scope of the light map
with a weighted average of the deposits:%
\begin{equation}\label{eq-WeightedAverageLightmap}
L_i(\vec{x}_1, \ldots \vec{x}_n, t) = \frac{\sum_{a=1}^n L_i(\vec{x_a}, t) E_a}{\sum_{a=1}^n E_a},
\end{equation}
where $E_a$ is the energy of deposit $a$.
This formulation depends on
a priori knowledge of the individual deposit energies,
which can be roughly determined for each deposit
by the measurement of the ionization signal.
This partitioning suffers from 
the anti-correlation between the
scintillation and ionization signals, 
so those values only allow us to 
approximate the true distribution of photons from the event.  
As a result
eq.~\ref{eq-LagrangeSystemOfEquations-c},
which guaranteed that the SS energy estimators would be unbiased,
can only approximately
guarantee an unbiased estimator for MS events.

\section{Statistical Characterization of EXO-200}
\label{sec-characterization}

The preceding section described de-noising
for a multichannel detector under generic
assumptions about the statistical properties
of signal pulses and noise.
In this section
these properties are quantified for the EXO-200 detector
using a combination of physical modeling
and calibration data to understand
the second-order statistics relevant
to de-noising.

\subsection{Additive Noise Correlations}
\label{sec-noise-corr}

\begin{figure}[t]
  \centering 
      \includegraphics[width=\columnwidth]
{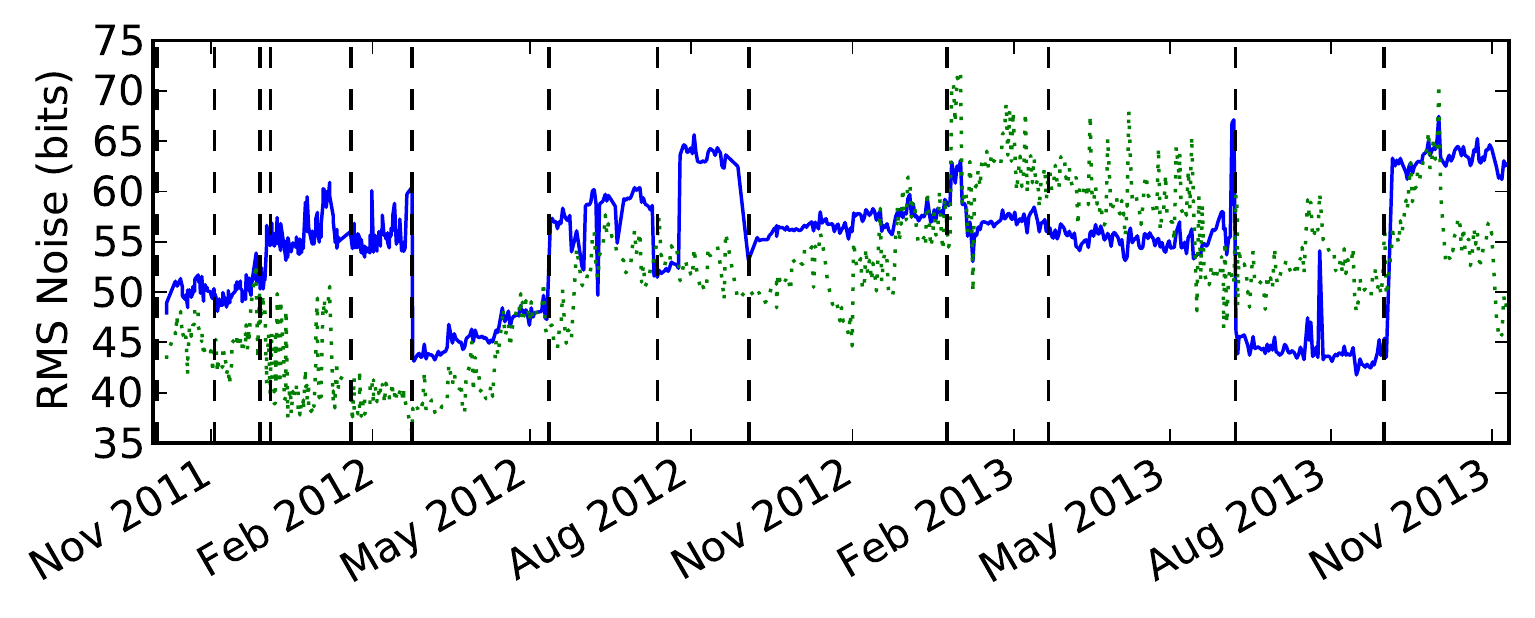}
\caption{%
Noise measured on the APD planes vs. date for
TPC 1 (solid blue) and TPC 2 (dotted green).
The dashed vertical lines indicate start dates 
of intervals of roughly stable APD noise behavior.}
  \label{fig-restimewindows}
\end{figure}
  
De-noising mitigates additive noise in two ways:
by exploiting correlations between channels
and by giving channels with worse signal-pulse-to-noise ratio lower weight.
Both can be accomplished by
specifying the second-order correlations between
noise waveforms for each Fourier frequency, $\Covar{N_i[f]}{N_j[f]}$.

The statistical properties of the noise are known to vary over time,
so noise measurements are gathered
during every physics run by a ``solicited'' trigger.
These solicited data frames are triggered independently of channel thresholds at a fixed \SI{0.1}{\Hz} rate to collect noise events for detector characterization.
Solicited-trigger data frames are almost
entirely free of signal pulses due to the \SI{\sim 0.1}{\Hz} rate of events collected by other triggers in EXO-200.
Any exceptional solicited trigger 
that is coincident with an energy deposit large enough
to exceed either the scintillation or ionization
detection thresholds is excluded prior to characterizing the noise.

Using a single set of noise properties for the entire data collection period
would 
not account for the  time dependence of the noise and
negatively impact the de-noised energy resolution. 
Conversely,
storing noise correlations for each physics run
is unnecessary.
In addition, it 
would lead to a partitioning
in which calibration runs rely
on nearby low-background data
for their noise calibrations,
whereas low-background runs could directly 
make use of their own waveforms 
to measure noise 
and might therefore benefit from a different level of accuracy in those inputs.
To alleviate such concerns,
the data collection period is binned into multiple time intervals,
each with fairly stable noise properties.

In figure~\ref{fig-restimewindows} dashed lines indicate the start of each such interval.
These boundaries are chosen at sudden changes
in channel behavior due to known environmental changes
(such as power outages, 
APD power cycling, 
temperature excursions, 
and changes to the electronics)
and other observed sudden changes in behavior.
Noise correlations are computed per interval,
averaging over all noise events that pass
cosmic-ray veto, timing coincidence,
and other criteria for rejecting spurious events.

\subsection{Scintillation Light Map}
\label{sec-lightmap}

The second input is the `light map' $L_i(\vec{x},t)$ 
which, 
as described in eq.~\ref{eq-lightmapdefinition},
characterizes the response of APD channel $i$,
for an event at position $\vec{x}$ 
occurring on date $t$.
The position dependence arises primarily from
geometric effects such as 
increased collection of the scintillation signal 
by the APDs nearest to the energy deposit.
In certain regions, 
e.g. near edges of the detector,
the majority of the light collection occurs on just a few APDs.
Individual APD channel gains 
have been observed to drift appreciably on time scales of months.
Sudden changes have been observed due to
several specific modifications of the detector readout electronics 
and occasionally due to environmental fluctuations.
The light map is determined empirically
by measuring the signal pulse amplitudes observed 
from predominantly mono-energetic events. 

The previous EXO-200 light map
(used in \cite{Albert:2013_2nuPRC})
characterized the summed APD response averaged over long periods.
A light map for each individual channel requires substantially larger calibration statistics,
per position-time bin,
than the previous form because of the larger statistical fluctuations
in the signal pulses observed on individual channels.
Also,
individual APD channel gains drift less gradually
than the average gain of an APD plane,
and near the edges of the detector the light collection of an individual APD channel
has greater dependence on the origin of the photons than the light collection of an entire APD plane.
Thus, finer light map binning, in both time and space, is required.

The $L_i(\vec{x},t)$ are determined 
from $^{228}$Th calibration events associated with 
the prominent \SI{2615}{\keV} spectral line.
The entire EXO-200 $^{228}$Th calibration data set is used,
providing more than $10^8$ total detector triggers.
However, 
these statistics are insufficient
to directly generate the full four-dimensional per-channel
light map required for de-noising.
To proceed, 
APD channel responses are assumed separable:%
\begin{equation}\label{eq-lmap-S}
L_i(\vec{x},t) = h_i(\vec{x})S_i(t),
\end{equation}
where the position-dependent response $h_i(\vec{x})$
depends only on the light collection geometry and represents how likely it is for a photon emitted isotropically at $\vec{x}$ to collect on the channel,
which is constant in time,
while the time-dependent response $S_i(t)$ accounts for 
variations in the APD gain with time,
which is independent of the position of the event.
This separability
should be valid for a single APD,
and any deviations from full separability,
due to variations of the gains within each APD gang
or changes in detector geometry over time,
are assumed to be negligible.
\begin{figure}%
\centering%
\includegraphics[width = 0.49\columnwidth]{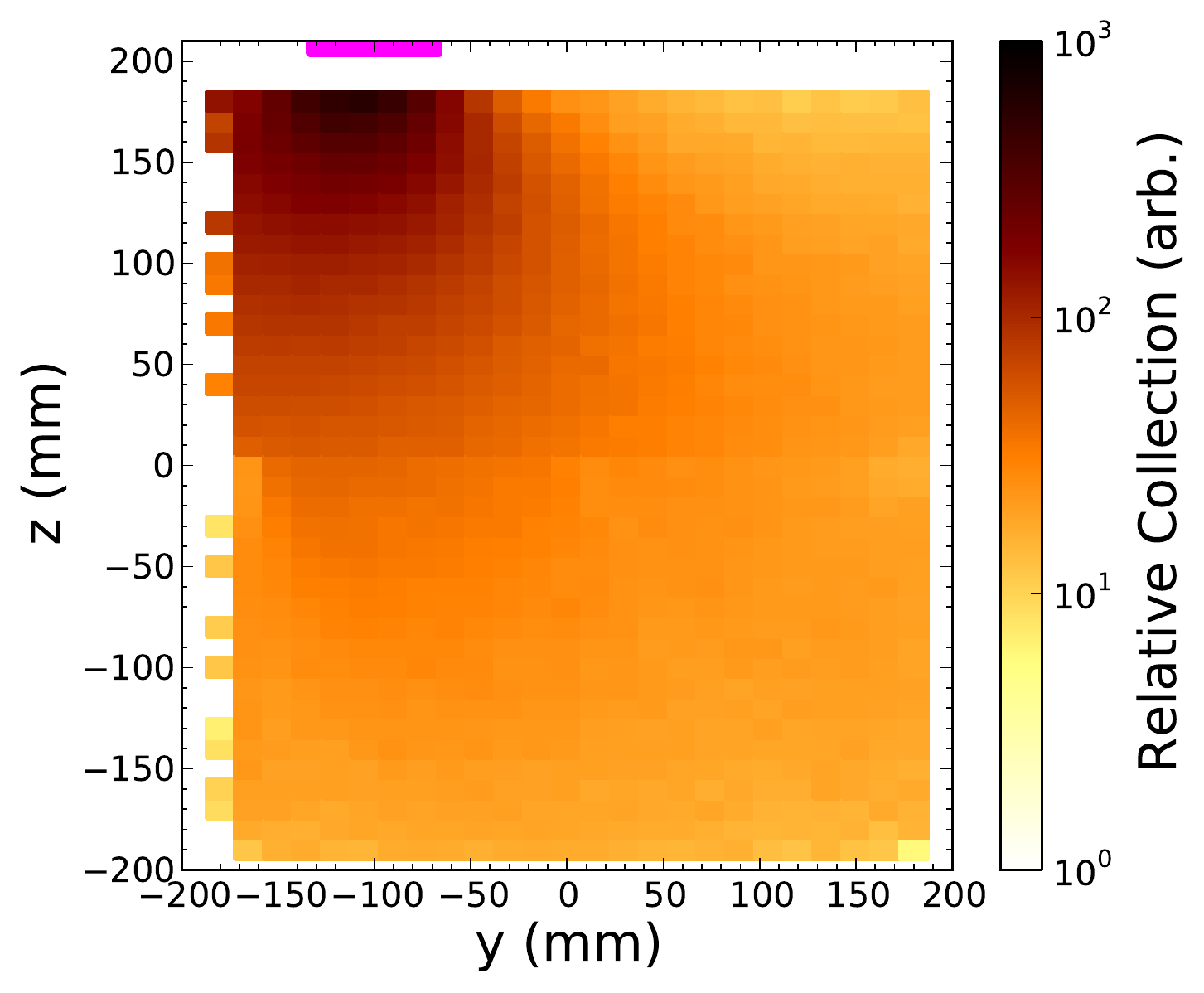}%
\includegraphics[width = 0.49\columnwidth]
{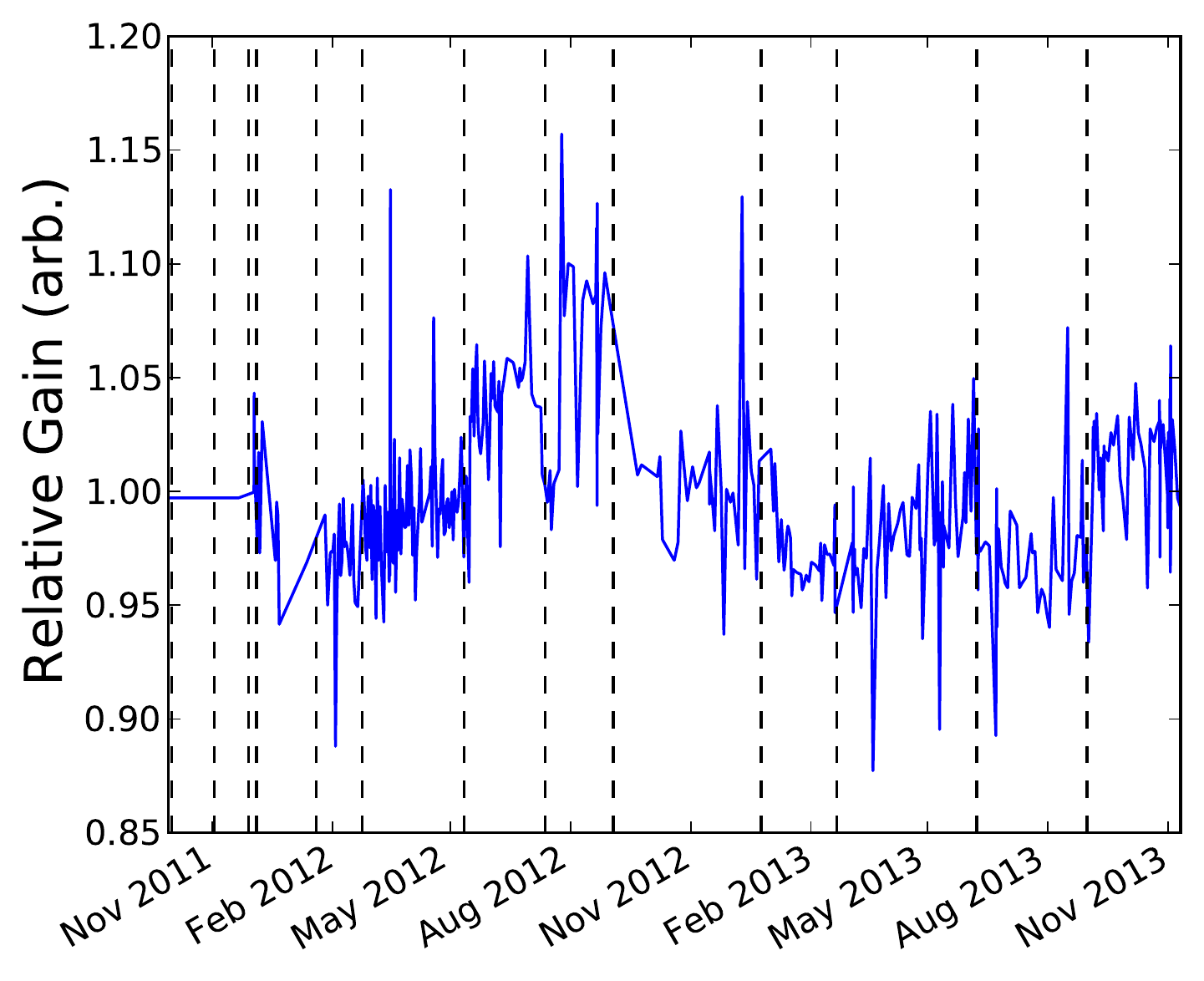}
    \caption{%
    (left) A slice of $h_{173}(\vec{x})$ along the length (\textsf{z}) and diameter (\textsf{y}) of the detector;
    collection is greatest near the APD gang associated with channel 173 (magenta box).
    Refer to figure~\ref{fig-detector} for the APD gang position.
    The cathode is at $\textsf{z}=0$.
    (right) $S_{173}(t)$, the relative gain change over time of channel 173. 
    For reference noise periods are indicated as in figure~\ref{fig-restimewindows},
    however the light map does not use such binning.}
    \label{fig-lmap}
\end{figure}

The light map is determined from the response 
to \SI{2615}{\kilo\eV} full-absorption SS events. 
The selection of such events is an iterative process 
since the light map itself must be used 
for the best estimate of event energy.
Events are first selected based on
the energy determined from charge information only.
An initial light map is then generated,
and used to determine an improved energy estimate for each event.
The final light map is then determined 
after re-selecting events associated with the \SI{2615}{\kilo\eV} peak.
Further iterations were not found 
to substantially alter the light map.

The position dependence of the light map $h_i(x)$
is determined by binning the data into
$1.5{\times}1.5{\times}1$~\si{\cubic\cm} voxels%
\footnote{%
The short dimension is parallel to the axis of the detector,
and is reconstructed by the time delay between light and charge collection.}
and linearly interpolating
between voxel centers.
The time dependence of the light map $S_i(t)$
is measured for each $^{228}$Th calibration
with sufficiently high statistics
and linearly interpolated between them.
Figure~\ref{fig-lmap} shows the 
spatial and temporal dependence of
the light map for a typical APD gang.

\subsection{Scintillation Noise Model}
\label{subsec-ScintillationNoiseModel}

The third input to de-noising characterizes fluctuations in signal pulse amplitudes.
The readout process from an energy deposit is inherently noisy,
since
the number of photons reaching each APD channel,
and the collection
and amplification efficiencies of the APDs vary 
from event to event.
In this section,
we discuss the expectation value of $\Covar{M_i}{M_j}$,
which reflects the variances and covariances
of the signal pulse amplitudes from a fixed energy deposit
(not considering additive noise).
Like the light map, 
these covariances
could be characterized empirically from additional calibration data,
but this is infeasible in practice 
due to limited statistics.
The following describes a physical model for
calculating the desired variances and covariances.
This model is found to be sufficiently accurate
for de-noising EXO-200 waveforms.

Let $P^{(0)}=E/w$
represent the number of photons produced by an event.
Simulations using NEST~\cite{Szydagis:2011tk}
of the ${}^{228}$Th $\gamma$ give
$w\approx{(\SI{2615}{\keV})}/{82000}$
at the bulk field in EXO-200.
The physical processes of
photon collection and conversion into a digitized signal pulse amplitude
introduce several sources of error.

Let $P^{(1)}_i$ be the number of photons collected
on APDs of channel $i$. 
$P^{(1)}_i$ is a random variable that depends on
the position of the event.%
\footnote{%
$P^{(1)}_i$ is assumed to be constant over time,
as discussed in section \ref{sec-lightmap}.}
Assuming its expectation value is proportional to $P^{(0)}$:%
\begin{equation}
\Expect{P^{(1)}_i} = h_i(\vec{x}) P^{(0)}.
\end{equation} 
This collection can be modeled by a multinomial distribution;
each photon is collected by at most one APD channel
and the distribution of photon destinations are independent
and identically distributed. This yields the second-order expectation value%
\begin{equation}
\Covar{P^{(1)}_i}{P^{(1)}_j} =
h_i(\vec{x})P^{(0)}\delta_{ij} - h_i\left(\vec{x}\right) h_j\left(\vec{x}\right) P^{(0)}.
\end{equation}

Considering the statistical properties of the APDs,
described in detail in \cite{Neilson:2009kf}, 
$P^{(1)}_i$ photons collected on an APD channel
produce an output of $P^{(2)}_i$ electrons
that has the following first- and second-order characteristics:%
\begin{subequations}\label{eq-P2characteristics}\begin{align}
\ExpectCond{P^{(2)}_i}{P^{(1)}_i}
&=  G^{\rm D}_i(t) G^{\rm Si} P^{(1)}_i\\
\CovarCond{P^{(2)}_i}{P^{(2)}_j}{P^{(1)}_i, P^{(1)}_j}
&= \left[\left[G^{\rm D}_i(t)\right]^2 F^{\rm Si}
+ \sigma^2_{G^{\rm D}_i}(t)
+\left[G^{\rm Si}P^{(1)}_i + F^{\rm Si}\right]\sigma^2_{\rm NU}\right]G^{\rm Si}P^{(1)}_i \delta_{ij}
\end{align}\end{subequations}
where:
\begin{itemize}
\item[$G^{\rm D}_i(t)$] 
is the time-dependent mean internal gain
of the APDs associated with channel $i$. This is directly proportional to the $S_{i}(t)$
measured as described in section~\ref{sec-lightmap}.
\item[$G^{\rm Si}$] 
is the mean number of electron-hole pairs generated 
in the active layer of an APD, $\sim$1.7. 
Incident \SI{7}{\eV} Xe scintillation photons require $\sim$\SI{4}{\eV}
per pair created
in silicon
at room temperature~\cite{Scholze:2000-208};
at \SI{167}{\K}, 
there is a 1.6\% increase due to temperature dependence~\cite{Lowe:2007367}, 
but this is less than the error on the measurement.
\item[$F^{\rm Si}$] 
is the Fano factor of silicon,
$0.118\pm 0.004$ \cite{Lowe:2007367}.
\item[$\sigma^2_{G_i^{\rm D}}(t)$] 
quantifies the variance of the internal gain experienced
by electron-hole pairs due to
the physical avalanche process,
$\sigma^2_{G_i^{\rm D}}(t) {\approx \left[G^{\rm D}_i(t)\right]^2}$ \cite{Neilson:2009kf}.
\item[$\sigma^2_{\rm NU}$] 
quantifies variance of the mean internal gain 
due to the non-uniformity of the APD channel response
(e.g. location of amplification within an APD,
distribution of collection on APDs within a gang,
angle of incidence),
which is not alleviated by higher photon statistics. 
Due to a lack of available data,
we fix $\sigma^2_{\rm NU}=0$,
but retain it symbolically for reference.
\end{itemize}

Finally, 
the current output from the APD channels is amplified
by standard electronics and digitized. 
This process, assumed to be noise-free, 
multiplies a current pulse of $P^{(2)}_i$ electrons 
by a time-dependent factor $G^{\rm E}_i(t)$,
producing a digitized pulse of magnitude $M_i$:%
\begin{equation}
M_i = G^{\rm E}_i(t) P^{(2)}_i.
\end{equation}

Combining these processes,
with $G_i(t) \coloneqq G^{\rm D}_i(t) G^{\rm E}_i(t) G^{\rm Si}$,
we arrive at the expectation values
\begin{subequations}\begin{align}
\Expect{M_i} ={}&
G_i(t)h_i(\vec{x})E/w,
\text{\quad and the covariances}\\
\Covar{M_i}{M_j} ={}&
\left(\begin{multlined}\label{eq-PulseCorrelationDetailedForm}
\left[1 +\frac{F^{\rm Si}}{G^{\rm Si}}\right]G_i(t)\left[1+{\frac{\sigma^2_{\rm NU}}{\left(G^D_i(t)\right)^2}}\right] \\
{}+\frac{G^{\rm E}_i(t)}{G^{\rm D}_i(t)}\sigma^2_{G_i^{\rm D}} + L_i(\vec{x}, t)\left[E-w\right]{\frac{\sigma^2_{\rm NU}}{\left(G^D_i(t)\right)^2}}
\end{multlined}\right)
L_i(\vec{x}, t) E \delta_{ij} - L_i(\vec{x}, t)L_j(\vec{x}, t) Ew.
\end{align}\end{subequations}
Note that 
eq.~\ref{eq-lightmapdefinition} and eq.~\ref{eq-lmap-S}
give $L_i(\vec{x}, t)=G_i(t)h_i(\vec{x})/w$
and $G_i(t)/w = S_i(t)$.

An interesting feature of
eq.~\ref{eq-PulseCorrelationDetailedForm} is its
dependence on
the energy $E$, which is our goal to estimate, 
reflecting the fact that shot noise (unlike additive noise)
is energy-dependent and that de-noising may prefer different
noise trade-offs depending on the energy of an event.
At first,
this would seem to indicate a circular dependency.
However,
charge-only data can be used as an independent estimate of
energy to cue de-noising with an appropriate estimate
of $\Covar{M_i}{M_j}$.
The estimator is still guaranteed to be unbiased,
and any discrepancy in $\Covar{M_i}{M_j}$ will
only make the estimate less accurate.
If this approach is inadequate,
it is also possible to iteratively refine the estimate
of $\Covar{M_i}{M_j}$ based on the de-noised estimate of $E$,
and repeat the de-noising technique.

\section{Resolution Improvement}
\label{sec-res-improve}

\begin{figure}[t]
	\centering
    \includegraphics[width = \columnwidth]{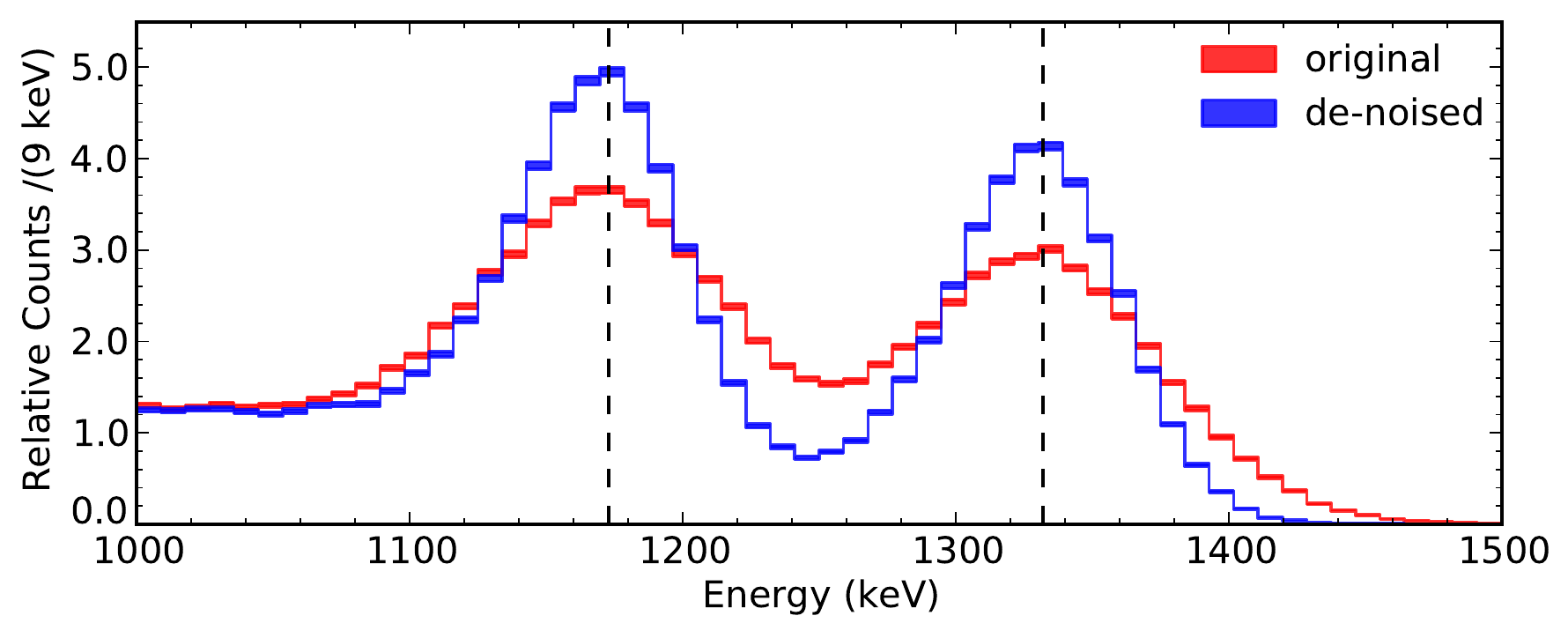}
    \caption{%
${}^{60}$Co source calibration data compared for original (red) and de-noised (blue) data. 
The bands show the statistical errors on the counts of SS events in each \SI{9}{\keV} energy bin.
The two gamma lines (1173 and \SI{1332}{\kilo\eV}) are indicated by dashed vertical lines.}
    \label{fig-co-data}
\end{figure}

EXO-200 determines its energy estimate for an event
through a linear combination of its scintillation and ionization energy estimates~\cite{Albert:2013_2nuPRC}.
De-noising improves the scintillation energy estimate.
This directly improves this combined energy estimate,
and also allows for an increase in the weight of the scintillation channel in calculating the combination, 
which again improves the estimate.
Here we discuss some ways in which de-noising has improved this combined energy estimate.
A comparison of ${}^{60}$Co source data with and without de-noising in figure~\ref{fig-co-data} shows this improvement.

\subsection{Time Independence}

\begin{figure}[t]
	\centering
    \includegraphics[width = \columnwidth]{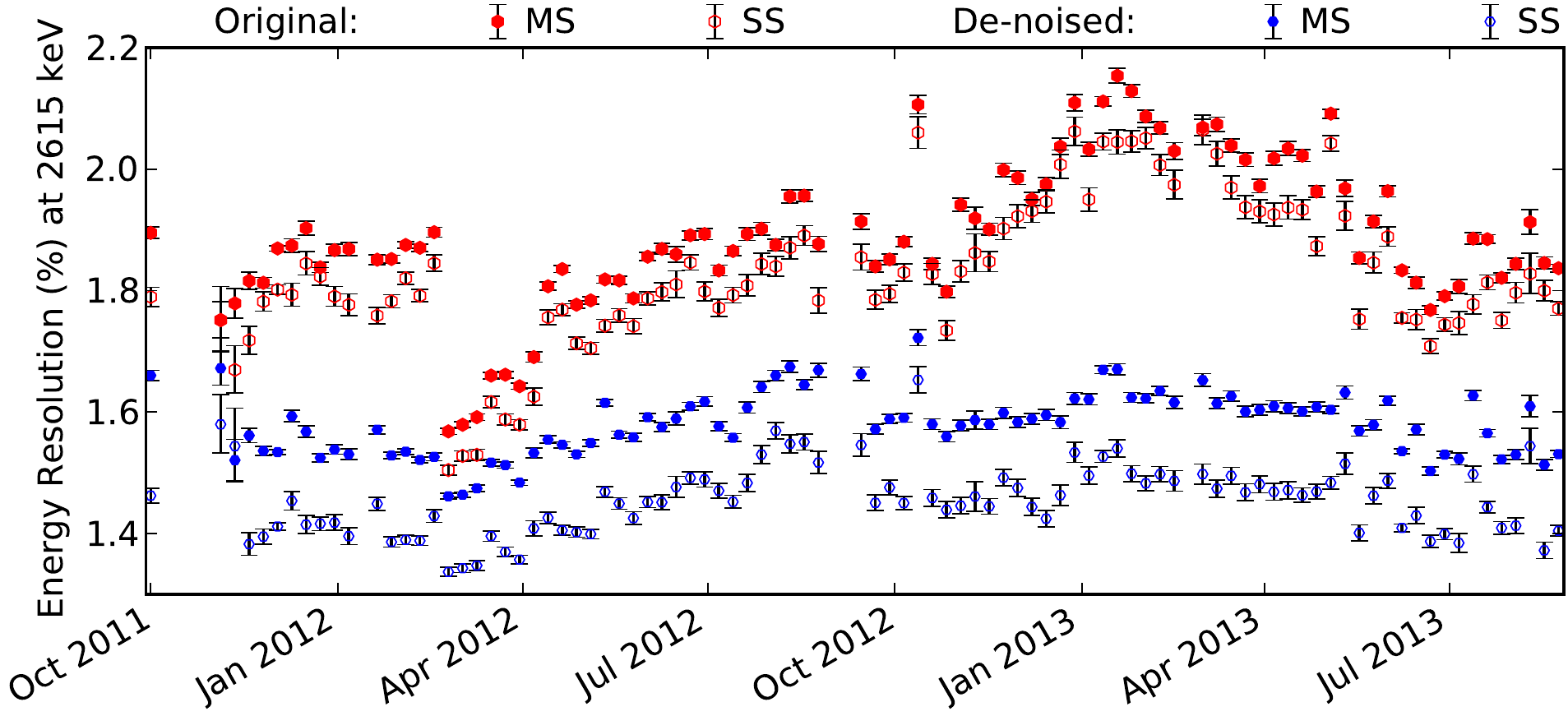}
    \caption{%
Original (red) and de-noised (blue) energy resolution 
at the ${}^{208}$Tl peak (\SI{2615}{\keV}) 
from ${}^{228}$Th calibrations
for multisite (MS, filled) and single site (SS, open) events.}
    \label{fig-restime}
\end{figure}

Recall (cf. figure~\ref{fig-restimewindows}) that 
the noise on the APD planes varied considerably over time.
Prior to de-noising, 
this resulted in a pronounced time dependence of the energy resolution.
De-noising removes much of this time dependence in two main ways.
First by heavily weighting few channels in each event
and leveraging noise correlations between channels,
the overall contribution of the additive noise is reduced 
(and the time dependence it causes). 
Second as the weightings are performed per event, 
de-noising dynamically makes the trade-off between additive noise and photon statistics.
Thus the more additive noise increases,
the more de-noising works to suppress its impact.

In determining the total effective detector energy resolution,
the weekly total energy resolution is first measured for the ${}^{228}$Th source calibrations by combining the ionization and scintillation responses.
These weekly resolution measurements 
are shown in figure~\ref{fig-restime},
where the total SS and MS energy resolution 
of the ${}^{208}$Tl full-absorption peak (\SI{2615}{\keV} $\gamma$) 
without (original) and with de-noising applied are compared.

\subsection{Improved Energy Spectra}

\begin{figure}[t]
    \centering
    \includegraphics[width=\columnwidth]{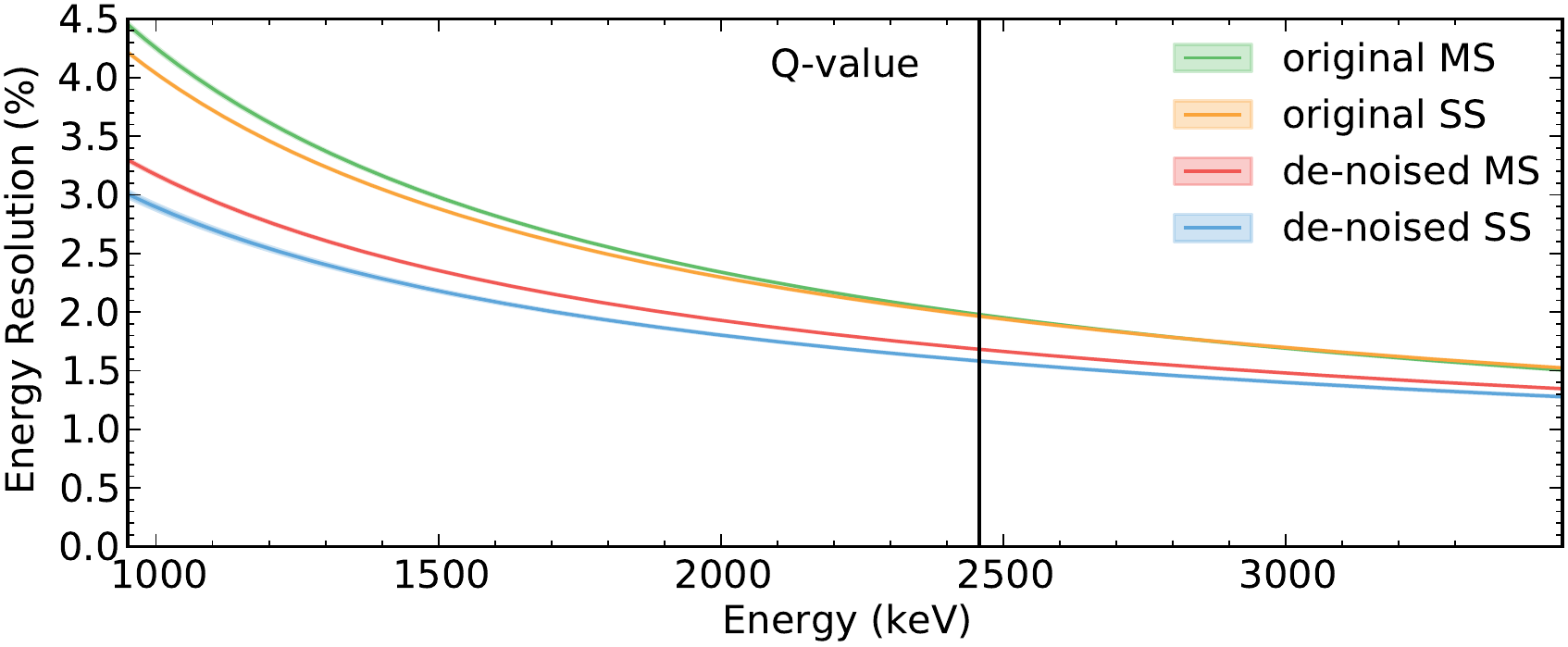}
    \caption{%
Relative resolution compared between 
original and de-noised,
multisite (MS) and single site (SS) data, 
with bands indicating the fit error propagated from the parameter covariance matrix.
}
	\label{fig-res}
\end{figure}
\begin{table}[t]%
\caption{%
Resolution function parameters and approximate errors
determined from fits to the calibration data.
The resolution evaluated at the \znbb Q-value is shown at right.
}%
\begin{center}%
\begin{tabular}{|rr|ccc|c|}%
\hline
& & & & & \\[-2ex] 
& &  $\sigma_{\rm A}$ (keV) & $b$ ($\sqrt{\rm keV}$) & $c$ (1e-3) & $\sigma\left(Q_{\beta\beta}\right)/Q_{\beta\beta}$ (\%) \\
\hline
\multirow{2}{*}{\small original} & MS &
$37.75\pm0.31$&$0.6187\pm0.0080$&$0.0\pm1.5$&$1.9790\pm 0.0027$\\
 		 & SS &
$33.85\pm0.06$&$0.6942\pm0.0049$&$0.2\pm5.9$&$1.9641\pm 0.0039$\\
\hline
\multirow{2}{*}{\small de-noised} & MS &
$22.86\pm0.23$&$0.6949\pm0.0035$&$0.0\pm1.4$&$1.6822\pm 0.0022$\\ 
		 & SS &
$19.40\pm0.69$&$0.6798\pm0.0088$&$0.0\pm6.7$&$1.5820\pm 0.0044$\\
\hline
\end{tabular}\end{center}%
\label{tab-res-par}%
\end{table}

The effective resolution takes the form 
$\sigma^2(E) {={}} \sigma^{2}_{\rm A} + b^2E + c^2E^2$,
where $\sigma_{\rm A}$ accounts for additive noise
that is dominated by APD electronics noise contributions, 
and
$b^2E$ and $c^2E^2$ account for statistical fluctuations and position- and time-dependent broadening.
To determine the resolution,
calibration data are fit to simulated energy spectra
produced using the Geant4-based EXO-200 Monte Carlo simulation software
(see \cite{Albert:2013_2nuPRC}).
${}^{137}$Cs, ${}^{60}$Co, ${}^{226}$Ra, and ${}^{228}$Th 
sources are used for this calibration data.
These sources provide prominent spectral features
throughout the 1--\SI{3}{\mega\eV} range shown in figure~\ref{fig-res}.
Table~\ref{tab-res-par} compares the fitted parameters
for the de-noised and original data.
Parameter errors were determined 
by inverting the matrix of second order derivatives
around the best fit values 
and are also reported in Table~\ref{tab-res-par}.
This method provides only an approximation of the parameter errors
due to the presence of strong correlations 
and the best fit value for parameter $c$ occurring near a physical boundary.
Errors on the resolution at the \znbb Q-value were propagated using the full covariance matrix from the fit.

Before de-noising, 
the effective resolution for SS (MS) at the \znbb Q-value is
\mbox{$[1.9641\pm 0.0039]$\%} (%
$[1.9790\pm 0.0027]$\%).
After de-noising, 
the effective resolution for SS (MS) at the \znbb Q-value is
$[1.5820\pm 0.0044]$\% (%
$[1.6822\pm 0.0022]$\%).
These fits, 
performed with improved Monte Carlo
which considers the full spectral shape of the sources rather than selected peaks,
are consistent with those reported in~\cite{Albert:2014awa}.
Reducing the correlated noise on the scintillation channel with de-noising 
has decreased $\sigma_{\rm A}$ by one third.
The other parameters are not significantly impacted by de-noising.
$\sigma_{\rm A}$ has the greatest effect on the relative resolution at lower energies.

\section{Future Work}

The current de-noising technique is most developed
for SS events and not optimized for MS events.
Future work will focus on exploring alternative approaches to MS
events which are identified in this section.

The current approach to MS events depends on an approximate extension
of the light map (eq.~\ref{eq-WeightedAverageLightmap}).
This approximation could generate a biased energy estimator.
This bias can be eliminated by replacing the SS bias-free constraint
of eq.~\ref{eq-LagrangeSystemOfEquations-c} with a stronger
MS version:%
\begin{equation}
\sum_j \left(\sum_{g} A_{j}[g] Y_{j}^{\rm R}[g] + B_{j}[g] Y_{j}^{\rm I}[g]\right) L_j\left(\vec{x}_b, t\right) 
= 1\text{\quad for all $b$,}
\end{equation}
where $b$ indexes deposition sites.
Such stronger constraints will increase the
expected estimator variance,
but the benefit is that the estimator
is guaranteed to be bias-free regardless of
the distribution of energy between deposits or
any error
in the approximate light map extension of eq.~\ref{eq-WeightedAverageLightmap}.

Beyond total scintillation energy,
it is also possible to define
estimators for the scintillation energies of individual deposits.
This approach is similar to the technique for multievent frames
described in section~\ref{sec-complex-events},
in which eq.~\ref{eq-LagrangeSystemOfEquations-MultiEventConstraint} is
used to place nulls between deposits
rather than events.
However,
the problem is more difficult because
MS event deposits can only be differentiated
by the positions of their deposits
by taking advantage 
of the detailed light map described in section~\ref{sec-lightmap}
whereas multievent frames could also use differences in scintillation timing.

The scintillation-only energy resolution of each deposition site
so obtained may be poor
because they can only be distinguished by 
the pattern of their photon deposits across channels.
Nonetheless, any anti-correlated
deposit energy able to take advantage of ionization and scintillation anti-correlation
would be an improvement over using the charge energy alone,
as current analyses do.
In a \znbb analysis this information would primarily serve to constrain background rates,
which are already well-constrained by existing physics analyses,
so the expected benefit is marginal. 
However
this improved energy resolution of individual deposition sites 
would improve sensitivity 
in searches utilizing individual interactions within an MS event
such as the de-excitation $\gamma$s produced in
\tnbb decays to excited states of
the $^{136}$Ba daughter nucleus (see \cite{Albert:2015ekt}).

The ability to assign scintillation energy to 
deposition sites within an event
would also be useful to better understand the detector
and fit backgrounds.  
In particular, 
differences between the light yield from $\gamma$s and $\beta$s
could be studied using e$^+$--\;e$^{-}$ pair production
from $\gamma$ calibration sources.
Also,
the ability to use MS events for light map construction would reduce 
the impact of source calibration on detector livetime.
Finally, 
the light map may be further improved 
in such a case of limited statistics
by moving to an unbinned technique.

\section{Conclusions}

We have demonstrated a technique, de-noising, 
to create an optimal energy estimator that 
is robust against correlated noise for simultaneous signals on all channels.
After applying this technique to existing EXO-200 data,
the resolution at the \znbb Q-value of \otsx was improved from 
\mbox{$\left[1.9641\pm 0.0039\right]$\%} to 
\mbox{$\left[1.5820\pm 0.0044\right]$\%}
for point like charge depositions such as \znbb events.
Future development of this technique may allow the 
scintillation energy of each interaction to be determined
for events depositing energy at multiple locations in the detector.  
This technique can also be generalized to other
experiments that measure signals using many readout channels 
that suffer from correlated noise components
or substantial variation in time or space 
of the response measured on each channel.

\acknowledgments{%
EXO-200 is supported by DOE and NSF in the United States,
NSERC in Canada, 
SNF in Switzerland, 
IBS in Korea,
RFBR-14-22-03028 in Russia,
DFG Cluster of Excellence ``Universe'' in Germany,
and CAS-IHEP Fund and ISTCP (2015DFG02000) in China. 
EXO-200 data analysis and simulation uses resources 
of the National Energy Research Scientific Computing Center (NERSC),
which is supported by the Office of Science of the U.S. Department of Energy under Contract No. DE-AC02-05CH11231.
We gratefully acknowledge
the KARMEN collaboration for supplying the cosmic-ray veto detectors, 
and the WIPP for their hospitality.%
}

\bibliographystyle{JHEP}
\urlstyle{rm} 
\bibliography{references}

\end{document}